\documentclass[runningheads]{llncs}
\usepackage{graphicx}
\usepackage{cite}
\usepackage{url}
\usepackage[bottom]{footmisc}
\usepackage [autostyle]{csquotes}    
\MakeOuterQuote{"}

\begin{document}
\title{Border Control and Use of Biometrics: Reasons Why the Right to Privacy can not be Absolute\thanks{This work is carried out as part of the EU-funded project SMart mobILity at the European land borders (SMILE) (Project ID: 740931), [H2020-DS-2016-2017] SEC-14-BES- 2016 towards reducing the cost of technologies in land border security applications.}}
\titlerunning{Border Control and Use of Biometrics}
%
\author{Mohamed Abomhara\inst{1}\and
Sule Yildirim Yayilgan \inst{1}\and
Marina Shalaginova\inst{1}\and Zoltán Székely\inst{2}}
\authorrunning{Mohamed Abomhara et al.}
%
\institute{Department of Information Security and Communication Technology, Norwegian University of Science and Technology, Gjøvik, Norway \\
\email{\{mohamed.abomhara, sule.yildirim, marina.shalaginova\}@ntnu.no} \and
National University of Public Service, Faculty of Law Enforcement, Hungary \\
\email{dr.szekely.zoltan@gmail.com}}
\maketitle              
\begin{abstract}
This paper discusses concerns pertaining to the absoluteness of the right to privacy regarding the use of biometric data for border control.  The discussion explains why privacy cannot be absolute from different points of view, including privacy versus national security, privacy properties conflicting with border risk analysis, and Privacy by Design (PbD) and engineering design challenges.

\end{abstract}

\section{Introduction}

Biometric technologies are automated methods of recognizing and verifying the identity of individuals based on physiological or behavioral attributes~\cite{bhatia2013biometrics}. They are used progressively more and are highly adopted at European borders~\cite{BiometricsKey2018,eu-LISA2018}. Strengthening border security, improving border crossing efficiency and facilitating effective migration control and enforcement are among the main grounds for utilizing biometric technologies~\cite{ICAO2017}. Despite the many advantages of biometrics, there are some limitations. The integration of biometric information systems employed for border control leads to increased surveillance that involves collecting and storing personal data such as fingerprints as individuals cross borders, apply for visas or request asylum. According to the General Data Protection Regulation (GDPR)~\cite{voigt2017eu}, personal information falls in a number of general categories, such as identity number and financial information (Article 4(1) GDPR) as well as special categories, for instance, biometric data, sexual orientation, medical information, personal activities, etc. (Article 9(1) GDPR). Such information is a valuable asset because it is crucial for all individuals to be able to keep it to themselves. On one hand, biometric technology has been proven to be cost-effective in enhancing border security, detecting fraud, helping improve border crossing efficiency as well as enabling effective migration control and enforcement~\cite{Abomhara2019}. On the other hand, biometric technology has serious impacts on privacy and data protection~\cite{liu2013bio}.

Major concerns with the use of biometric technology relate to individuals' privacy reduction and the immutable link between biometric traits and persistent information storage about a person. The tight link between personal information and biometrics can have both positive and negative consequences for individuals' privacy. Recent research~\cite{dantcheva2016else} explores the possibility of extracting supplementary information from primary biometric traits, e.g. face, fingerprints and iris. These traits denote personal attributes like gender, age, ethnicity, hair color, height, weight and so on. However, a breach (unauthorized acquisition, access, use or disclosure) of such confidential information would violate the principle of the right to privacy for those consenting to cross a border, even if the breach involved innocuous information that would not result in any social, economic, legal, or any other harm.

The right to privacy is described as the right of the individual to be let alone and to decide how, when, with whom and to what degree their personal data should be shared and communicated~\cite{kizza2013ethical,zeadally2015privacy}. However, unlike absolute fundamental rights as "the right to life or the prohibition of torture and inhuman or degrading treatment or the right to be free from slavery," which admit of no restriction (judgment of Court of Justice of the European Union (CJEU) of 12 June 2003, Schmidberger, Case C-112/00, paragraph 80~\cite{case122}), the right to privacy is not an absolute right and hence can be limited by law~\cite{jonsson2016right,mironenko2011body} – for example in time of public emergency that threatens the life of a nation. The arguments made in this paper pertain to why privacy cannot be absolute from different points of view: (1) privacy versus national security; (2) privacy properties conflicting with border risk analysis; and (3) Privacy by Design (PbD) and engineering design challenges.

The remaining part of this paper is organized as follows. Section \ref{Background} briefly discusses the use of biometrics for border control and the right to privacy principle. Section \ref{privacy} investigates and argues why privacy cannot be absolute. Section \ref{Con} concludes the paper.

\section{Background}
\label{Background}

\subsection{The use of biometrics for border control}
\label{border}

The dramatic advances in biometric technologies have opened doors to unprecedented opportunities in the field of border control. The challenge of border security is to identify with assurance who is crossing the border and decide if the person is authorized to cross or not. Unassisted by technology, a border staff member cannot maintain this degree of swift, assured identification. Border authorities are equipped with biometric technologies to facilitate more efficient checks at borders and contribute to preventing and combating illegal migration, etc.~\cite{ICAO2017}.  Biometrics enable accurate identification since each person has their own unique physical characteristics. Moreover, the use of multimodal biometrics~\cite{parkavi2017multimodal,khoo2018multimodal} offers even better results with higher accuracy by combining several biometrics~\cite{lumini2017overview}. Multimodal biometrics-integrated border management benefits all stakeholders, including governments concerned with securing national territory, immigration authorities managing controls at ever more crowded borders, and simply travelers who want to enjoy the journey to their destination~\cite{BiometricsKey2018}.

European Union (EU)-wide border biometric information management systems~\cite{eu-LISA2018}, including the European Asylum Dactyloscopy Database (EURODAC), Visa Information System (VIS) and Second-generation Schengen Information System (SIS II) have an increasingly important role in the identity establishment process by storing biographic and biometric data of third-country nationals~\cite{robinson2014information,boehm2011information}. In addition, the centralized Entry/Exit System (EES) of border control is expected to be fully implemented by 2020 in compliance with Regulations (EU) 2017/2225~\cite{Regulations20172225} and (EU) 2017/2226~\cite{Regulations20172226}. Interoperability will allow the EU information systems to complement each other, help facilitate the correct identification of persons, contribute to fighting identity fraud and ease information sharing. Accordingly, the SMart mobILity at the European land borders (SMILE)\footnote[1]{http://smile-h2020.eu/smile/} interoperability with other border information systems is greatly promising in terms of enhancing the speed, efficiency and flow of border crossing mobility as well as border security.

As a border control tool, SMILE encourages the propensity to collect, use and process sensitive biometrics like fingerprints, face and iris data, etc. to improve traveler flow and boost border security. On the one hand, SMILE technologies are intended to enhance security levels and make the traveler identification and authentication procedures easy, fast and convenient. On the other hand, similar to other biometric technologies, SMILE technologies have raised new threats to fundamental rights, data protection and privacy.

\subsection{Right to privacy}

Every individual has the right to the privacy protection of their personal information when it is collected and shared. Generally, legal documents on data protection and privacy such as GDPR~\cite{voigt2017eu} refer to personal information protection throughout all steps from collection to storage and dissemination. Moreover, provisions of the European Convention of Human Rights (ECHR) constitute the basis for challenging inequitable decisions of public authorities~\cite{jonsson2016right,warren2019right,ccinar2019right}. Article 8 ECHR ensures everyone's right to have their private and family life, home and correspondence respected without public authority interference. The aforementioned right is mirrored in Article 7 of the Charter of Fundamental Rights (CFR). Moreover, Article 8 of the Charter enshrines the right of everyone to the protection of their personal data. Individuals (data subjects) have the right to exercise control over their personal data (Article 12-23 GDPR~\cite{voigt2017eu}). Protection of privacy is frequently seen as a line drawn for how far society can intrude into a person’s private affairs~\cite{mironenko2011body}.

However, these rights can be overruled if a legal basis is laid down for collecting, processing, storing or retaining personal data to achieve a legitimate goal~\cite{bonnici2014exploring}. According to Article 52(1) CFR, "subject to the principle of proportionality, limitations of rights to privacy may be made only if they are necessary and genuinely meet objectives of general interest recognized by the Union or the need to protect the rights and freedoms of others." Moreover, Recital 4 GDPR acknowledges that the right to data protection (as well as the right to privacy) is not an absolute right. Ruling the Eifert case~\cite{caseEifert}, the CJEU holds that "the right to the protection of personal data is not an absolute right, but must be considered in relation to its function in society." Furthermore, the rights to privacy must be proportionately balanced with other fundamental rights. According to the Opinion  of  Advocate  General  Jääskinen  in  the  Google  Spain  Case~\cite{caseGoogle},  the fundamental rights to privacy and to the protection of personal data are not absolute and may be limited provided there is a justification acceptable in view of the conditions set out in Article 52(1) of the Charter.

\section{Viewpoints of privacy: Why it can not be absolute}
\label{privacy}

This section discusses the contradictory interests of privacy versus national security, privacy properties conflicting with border risk analysis and Privacy by Design (PbD) and engineering design challenges.

\subsection{Privacy versus national security}
\label{PVS1}

The Schengen Borders Code (SBC) (Regulation (EU) 2016/399)~\cite{Regulations2016399} and its amendment (Regulation (EU) 2017/458)~\cite{Regulations2017458} set out the rules governing the movement of people across EU's internal and external borders. Internal borders means (a) the common land borders, including river and lake borders, of the Member States; (b) the airports of the Member States for internal flights; (c) sea, river and lake ports of the Member States for regular internal ferry connections. External borders means the Member States' land borders, including river and lake borders, sea borders and their airports, river ports, sea ports and lake ports, provided that they are not internal borders. SBC also defines the rules for the border checks of persons crossing external Schengen borders (border checks on persons). Cross-border movement at external borders shall be subject to minimum and thorough checks by border guards (Article 8 of Regulation (EU) 2016/399). The main objectives of the minimum and thorough checks are to ensure that the persons in question do not represent a threat to public order, internal security or public health, and to improve the security of the EU Member States and their citizens. Therefore, the key issue in question is how to achieve a trade-off between border check requirements (Regulation (EU) 2016/399) to ensure border security and meeting the need for flexible border crossing without compromising individuals' privacy.

According to a survey and interviews carried out for the SMILE project with land border guards (refer to the SMILE public deliverables), the majority of guards claim that the use of biometric authentication and verification at land borders is justified and necessary to improve border security measures and better protect the public interest. Biometrics are believed to help improve the accuracy of traveler identification and verification, meaning the ability to correctly recognize a genuine person and reject an imposter. Moreover, utilizing biometrics promotes the reduction of identity fraud (e.g., fake IDs and passports) as the identification and verification processes do not rely on the human agent. To the best of our knowledge, fraud reduction signifies accuracy increase. Therefore, using biometrics eliminates a considerable integrity threat that border guards face and benefits the authority responsible for border control. As a consequence, biometrics would result in a higher throughput of low-risk travelers without losing accuracy or integrity and allow human resources to focus on potentially higher-risk travelers.

The right of people to be respected for their private and family life, home, etc. (Article 8 of the ECHR) and also protected against unreasonable biometric searches (e.g., unreasonable biometric authentication and verification at the border) shall not be violated. For border authorities to be reasonable with using biometrics, traveler authentication and verification using biometrics must be as limited in its intrusiveness as it is consistent with satisfying the administrative need that justifies it~\cite{lind2015privacy}. It is important to argue on the one hand whether the introduction of biometrics for border control can be regarded as being in accordance with the law and if the biometrics satisfy a legitimate aim (public safety, crime prevention, etc.), proportionality and the necessity principles pursuant to Article 8 of ECHR. On the other hand, it is arguable whether breaching the right to privacy is also considered proportionate to the threats that biometric technologies are supposed to prevent. As for necessity and effectiveness, many discussions address whether biometrics actually add value to serving border control interests, including but not limited to, providing the safe, secure, efficient and unobtrusive, on-the-move security control of travelers, fighting terrorism and serious crime, and ensuring high internal security levels.

If national security has greater priority, should individual privacy yield? Besides, would it be ethically justifiable to sacrifice some privacy interests to achieve the highest national security gains possible? Moreover, if a traveler faces the dilemma of either providing biometrics or not being allowed to cross the border (in case no other alternative is provided), the person's right to freedom of movement may be restricted~\cite{mironenko2011body}. Our conclusion is not yet sufficiently clear to answer these questions. Even if it is accepted that using biometrics at borders is necessary and proportionate, serious concerns about whether the intrusion is in accordance with the law still remains a question. It may be said that conflicts arising between border security (Regulation (EU) 2016/399) interests and individual privacy cannot be weighed because they are not measurable by the same standards~\cite{leese2015privacy}. While acknowledging there could be legitimate aims for the invasion of privacy, the effectiveness of biometrics is still questionable. Therefore, a sensible trade-off is necessary between individual interests (individual privacy) and the legitimate concerns of the Member States such as preventing and investigating crime~\cite{valkenburg2015privacy,rung2017privacy}. An appropriate trade-off would involve retaining the benefits of biometric technologies to extend the border control ability to support high border security and reliability while maintaining individual privacy.

\subsection{Privacy properties conflicting with border risk analysis}
\label{PVS2}

Border risk analysis is a governance tool to normalize border and migration risks. It is based on an automated analysis of lager databases (SIS II, etc.) to extract useful information about people and their activities in order to identify behavioral patterns that may point to suspicious activity. Although border risk analysis is useful in decision-making, it can also lead to serious privacy concerns.

The key issue is in the contradictory interests of the principle of data minimization (Article 5 GDPR) that limits personal data collection, use and disclosure and the benefit of the capability to process personal data for performing border risk analysis. The main concerns include the unnecessary and unauthorized collection of biometric data for traveler identification and verification~\cite{campisi2013security,zeadally2015privacy}. For one, the data minimization principle and strong privacy properties (i.e., unlinkability, anonymity, undetectability)~\cite{deng2011privacy} restrict personal data collection and use for further analysis. However, border risk analysis requires the use of personal data for the investigation and/or monitoring of actions/activities of one or more individuals. In other words, the more personal data border authorities, for instance, can obtain about individuals, the better the risk prediction and thus overall risk analysis results will be. There has certainly been considerable progress in privacy preserving techniques for data analytics~\cite{rao2018privacy,saranya2015survey}. Nonetheless, even with such privacy-friendly techniques, border risk analysis is prone to privacy violations.

Another major concern is related to the potential for function/purpose creep. Purpose creep occurs when personal data is collected for one specific purpose and subsequently used for another unintended or unauthorized purpose without the user's consent. A famous example of a large-scale biometric function creep is related to the European Dactyloscopy (EURODAC) fingerprint database (Regulation (EC) 2725/2000). The original purpose of the EURODAC database was to compare fingerprints for the effective application of the Dublin convention (Regulation (EU) 603/2013)~\cite{zeadally2015privacy}. EURODAC enables EU countries to identify asylum applicants as well as illegal immigrants within the EU. However, soon after the database was established, other police and law enforcement agencies were also granted access. There are many other large-scale, centralized EU national and international databases, such as SIS II and VIS with the same risks. Similar concerns also arise in the case of border control risk analysis~\cite{Abomhara2019}. Hard and soft biometrics~\cite{dantcheva2011bag,zewail2004soft,abdelwhab2018survey} are likely to strengthen the potential for function creep due to the very sensitive nature of the data collected and the possibility to use centrally stored biometric data for purposes other than the original purpose (border crossing). Moreover, such databases offer more attractive targets for outsider attacks and insider misuse.

Therefore, the purpose specification principle (Article 5(1)(b) of GDPR), which is among the main principles of EU data protection legislation, has a key role. According to Article 5(1)(b) of GDPR, personal data must be collected for specified, explicit and legitimate purposes and not be further processed in a way incompatible with those purposes. However, GDPR provides exemptions in Article 23, which stipulates that Member States' laws may restrict the scope of the principles mentioned in Article 5 of GDPR when such restriction constitutes a necessary measure to safeguard national security and public security, to prevent, investigate, detect or prosecute criminal offences or execute criminal penalties, to protect the data subject or the rights and freedoms of others, etc. In our view, the indication of "a necessary measure" means that exemptions are restricted to specific investigations, case-by-case requests, and not to cases where personal data processing is systematic as foreseen by the use of biometrics for border control. As mentioned earlier, the processing of biometric data is questionable even when considering that an exemption might be applicable. Therefore, the problems of function creep and purpose misuse are not to be underestimated. Nonetheless, they can be curbed by stricter laws, particularly by limiting the use of specific biometric data for certain purposes. It can thus be concluded that the clarity of purpose regarding the intention of biometric data collection is paramount. It is important to be clear about the necessity for biometrics and how biometrics will help fulfill specified needs.

\subsection{Privacy by Design (PbD) and engineering design}

Privacy by design (PbD) is a policy measure that guides software developers to apply inherent solutions to achieve better privacy protection~\cite{hadar2018privacy}. For privacy to be embedded in the system development lifecycle and hence in organizational processes, system developers and policy makers must be ready to embrace and understand the domain~\cite{spiekermann2012challenges}.

Recent studies~\cite{sheth2014us,hadar2018privacy,senarath2018developers,oetzel2014systematic} reveal that most software developers lack formal knowledge and understanding of the concept of informational privacy. Besides, most have insufficient knowledge of how to develop privacy practices such as PbD~\cite{hadar2018privacy}. Software developers additionally find it difficult to understand privacy requirements by themselves~\cite{sheth2014us} and require significant effort to estimate privacy risks from a user perspective in order to relate privacy requirements to privacy techniques~\cite{oetzel2014systematic}. Moreover, software developers have trouble evaluating whether they have successfully embedded PbD strategies into the system design.

Privacy design frameworks serve as potential bridges between users, software developers and policy makers. Several studies like~\cite{kalloniatis2008addressing,gurses2011engineering,spiekermann2008engineering,hoepman2014privacy} discuss privacy design frameworks to assist software developers with addressing privacy during the system development process. However, to the best of our knowledge, it is still unclear how effective these design frameworks are and what the possible limitations for their utilization in everyday privacy engineering practices are. The key elements of PbD are intended to limit the collection, use and disclosure of personal data, to involve individuals in the data lifecycle, and to apply appropriate safeguards in a continuous manner~\cite{leese2015privacy}. This means separating personal identifiers, using pseudonyms and anonymization as well as deleting personal data when no longer needed~\cite{schaar2010privacy}. However, as argued by Leese~\cite{leese2015privacy}, such practices are undeniably suitable in economic and organizational contexts. But as discussed in Sections \ref{PVS1} and \ref{PVS2}, border checks and border risk analysis derive decisions exactly through the collection and processing of data, which could ultimately be connected to possible criminal activities, in order to control any risk. On the contrary, essentially PbD principles radically exclude the possibilities that come with advanced data analytics in border control contexts. Thus, the contradictory interests of PbD principles and the benefit of the ability to process biometrics data for border control cannot simply be resolved by technical means.

Thus, if PbD is ever to become a viable practice, a considerable change must be made to prepare the field for the wide implementation of this policy. Privacy implementation guidelines should be provided to help software developers and policy makers embed privacy into the system design. Moreover, an evaluation and demonstration of privacy assurance – as recommended by the ENISA\footnote[2]{European Union Agency for Network and Information Security (www.enisa.europa.eu).} guidelines for privacy and data protection~\cite{danezis2014privacy} -- is required to provide software developers with feedback to verify whether they have successfully followed the guidelines. This would reduce software developers' personal opinions on privacy practices and ease how privacy is embedded into the system. Moreover, the development method (Agile Software Development~\cite{abrahamsson2017agile}, Security Development Lifecycle~\cite{howard2006security,shostack2014threat}, etc.) used within the organization must be taken into account in order to apply the concepts of privacy throughout the entire system development process. This will enable development teams, policy makers, etc. to take appropriate measures in the relevant phases. Finally, upon design completion, the organization must adopt and monitor it throughout its lifetime.

Alongside the PbD issue is the Privacy by Default obligation. Under this obligation, data controllers must implement appropriate measures on both technical and organization levels to ensure that personal data collected is only used for specific purposes. Essentially, only the minimum amount of personal data required should be collected and stored, while data subjects should be allowed data accessibility and controllability. Ensuring privacy through every phase of the data lifecycle (collection, use, retention, storage, disposal or destruction) has also become crucial to avoiding legal liability, maintaining regulatory compliance, and so on. Therefore, integrating the Data Protection Impact Assessment (DPIA) with the Ethical Impact Assessment (EIA) and Privacy Impact Assessment (PIA) in the earlier stages of any system development would aid with the early identification of ethical and privacy problems and risks. Ideally, a full and detailed description of the processing along with its necessity and proportionality would help manage the risks to the rights and freedoms of natural persons resulting from personal data processing. Furthermore, taking PbD strategies into consideration should precede system design to ensure that ethical and privacy principles are taken into account. As a result, data controllers will be more able to comply with the legal requirements of data protection and demonstrate taking appropriate measures where DPIA is used to check compliance against data protection regulations.

\section{Conclusions}
\label{Con}

Border control systems raise the tendency to collect, use and process personal data (e.g. alphanumeric data like names and birth dates; biographic information; biometric data like fingerprints) to optimize and monitor the flow of people at land borders as well as enhance security and detect fraud. However, evidence demonstrates that personal data collection and processing pose several privacy challenges. Privacy is a fundamental human right in EU countries and is controlled by legislation that responds and adapts to data subjects' privacy needs. Moreover, the obligations of the Member States pertaining to personal data collection and processing along with the exchange of personal data among Member States are stated in various EU laws and regulations. It is therefore essential for border authorities to consider the legal consequences of developing and deploying biometric identification and authentication methods.

Personal data flows and ripples are in some ways difficult to predict. Despite all attempts to provide anonymity, biometric data still penetrates a person's physical, psychological and social identity. Biometric technology enables revealing personal information, such as gender, age, ethnicity and even critical health problems like diabetes, vision problems, Alzheimer's disease, etc. Members of particular groups including disabled, transgender and older people, religious groups, and others, can encounter additional negative effects on privacy. Although numerous proposals, recommendations and legal considerations are in place as safeguards, it is unclear how they can ultimately be put in practice. For example, even a well-conceived, general and sustainable data protection and privacy regulation like GDPR is strained by the effort to ensure superior effectiveness with respect to privacy.

It is quite obvious that biometric technology exposes travelers to significant loss of privacy and limitations of other rights and freedoms. This paper discussed concerns with the absoluteness of the right to privacy regarding the use of biometric data for border control. In accordance with Article 8(2) ECHR, "there shall be no interference by a public authority with the exercise of the right to privacy except such as is in accordance with the law." The exemption clause contains two conditions: (1) the necessity for a democratic society that should be proportional to the purposes of the law and (2) exceptions in accordance with the law. These conditions require a specific legal rule that authorizes the interference and sufficient access of individuals to the specific law, and that the law must be precisely formulated in order to ensure that individuals are capable of foreseeing the conditions of its applicability.

In future, the authors plan to investigate the effect of using biometrics for border control on individuals' privacy. Unfortunately, there is still too little knowledge about the real effects on individuals and a system's reputation when privacy breaches occur. This is because on the one hand, very little knowledge exists about the tangible and intangible benefits of personal data collection in EU information systems. On the other hand, it is not clear to what extent people have the right to choose what information about themselves to share and how to engage with border systems and devices such as biometric sensors and readers.


%
%
%
\bibliographystyle{splncs04}
\bibliography{mybibliography}

\end{document}